\documentclass[journal,comsoc]{IEEEtran}
\usepackage[T1]{fontenc}

\usepackage{cite}

\ifCLASSINFOpdf
  \usepackage[pdftex]{graphicx}
  \graphicspath{{../pdf/}{../jpeg/}}
  \DeclareGraphicsExtensions{.pdf,.jpeg,.png}
\else
\fi
\usepackage{amsmath}
\usepackage[cmintegrals]{newtxmath}
\usepackage[justification=centering]{caption}
\usepackage{multirow}
\usepackage{tabularx}
\usepackage{threeparttable}
\usepackage{makecell}
\usepackage{booktabs}  
\usepackage{multicol}
\usepackage{array}

\begin{document}
%
\title{Placement and Routing Optimization Problem for Service Function Chain: State of Art and Future Opportunities}

\author{\IEEEauthorblockN{Weihan Chen\IEEEauthorrefmark{1},
Xia Yin\IEEEauthorrefmark{2}, Zhiliang Wang\IEEEauthorrefmark{3} and
Xingang Shi \IEEEauthorrefmark{4}}\\
\IEEEauthorblockA{Department of Computer Science and Technology,
Tsinghua University,
Beijing\\
Email: \IEEEauthorrefmark{1}chenwh18@mails.tsinghua.edu.cn,
\IEEEauthorrefmark{2}yxia@tsinghua.edu.cn,
\IEEEauthorrefmark{3}wzl@cernet.edu.cn,
\IEEEauthorrefmark{4}shixg@cernet.edu.cn}}

%



\maketitle

\begin{abstract}
  Network Functions Virtualization (NFV) allows implantation of network functions to be independent of dedicated hardware devices. Any series of services can be represented by a service 
  function chain which contains a set of virtualized network functions in a specified order. From the perspective of network performance optimization, the challenges of deploying service 
  chain in network is twofold: 1) the location of placing virtualized network functions and resources allocation scheme; and 2) routing policy for traffic flow among different instances 
  of network function. This article introduces service function chain related optimization problems, summarizes the optimization motivation and mainstream algorithm of virtualized network 
  functions deployment and traffic routing. We hope it can help readers to learn about the current research progress and make further innovation in this field.  
\end{abstract}


%
\IEEEpeerreviewmaketitle

\section{Introduction}
%
%
%
%
Service Function Chain (SFC) \cite{halpern2015service} refers to connecting different network functions in specific sequence and providing corresponding service for users. 
The network functions in SFC are realized as different Virtualized Network Function (VNF). In actual network, SFC can be configured and adjusted according to different traffic demand. 
The configuration process involves two aspects: the placement of VNF and traffic steering among different VNFs. In terms of VNF placement, 
the network operators (or Internet Service Providers) need to select the location for VNF Instance (VNFI), which can run VNF and allocate the resource (CPU, memory, etc.) for each VNFI. 
And in terms of traffic steering (routing), the path used to transmit traffic through specific VNFs of SFC needs to be determined. 
Proper SFC configuration can be helpful for improving network performance and reducing operational cost.

In actual network environment, both users and network operators have their own performance requirements for network functions. 
For network operators, the requirements can be reducing VNF placement cost and improving resource utilization. And for common users, the requirements can 
be increasing network throughput and reducing traffic transmission delay. These performance requirements need to be satisfied by adopting appropriate SFC configuration 
(including VNF placement and traffic routing). However, different VNF placement and traffic routing schemes for SFC may affect network performance and operational cost. 
It is difficult to find optimal SFC configuration only depending on human experience. By modeling optimization problem for VNF placement and traffic routing and solving the problem, 
determining corresponding SFC configuration schemes and satisfying performance requirements can be easier.

During the modeling process, the placement and routing optimization problem can be considered independently or jointly. 
When treating VNF placement optimization problem independently, VNF deployment and operational cost is considered as the prior optimization objective, 
the cost may include minimizing placement cost (mentioned in \cite{ghaznavi2017distributed}), minimizing traffic switching cost among different VNFs (mentioned in \cite{luizelli2018optimizing}), 
etc. And the constraints of placement problem mainly focus on resource capacity constraints, which can be host CPU core number, link capacity or other network resources. 
In contrast, the optimization objective of routing problem tends to prioritize routing cost. It aims to find a path with least cost. The cost has many choices 
(such as financial cost, delay, QoS requirement, etc. mentioned in \cite{dwaraki2016adaptive}). Meanwhile, 
the main constraint of routing problem is that user traffic flow should pass through the services provided by the SFC in the specified order.

On the other hand, in order to achieve better network performance, the VNF placement problem and traffic routing problem can be considered jointly. 
The optimization objective can be the combination of placement and routing optimization objectives. The constraints are also similar with the VNF placement 
optimization problem constraints plus routing constraints. However, optimizing VNF placement and routing jointly may cause some conflict. Because lower placement cost means 
less VNFIs are deployed, which results in higher routing cost (some traffic may be routed to longer path in order to achieve necessary network functions). On the contrary, 
to realize lower routing cost, more VNFIs need to be deployed, which causes placement cost increasing. Hence, finding a trade-off solution for joint optimization problem is necessary.

Currently, there is a great deal of research focuses on placement and routing optimization problem for SFC. 
They use different methods to model the optimization problem and develop corresponding algorithms to solve the problem efficiently. 
The algorithms try to find optimal SFC configurations in order to provide better network services for users with lower related cost. In this survey, 
we mainly focus on summarizing existing research about VNF placement and traffic routing optimization problem for SFC configuration. First, we introduce existing solutions of independent 
VNF placement problem and traffic routing problem, and then the joint optimization problem of placement and routing will be discussed. Each kind of optimization problem is presented in detail. 
In addition, we also discuss the future opportunities for placement and routing method of SFC.

 

\section{Virtual Network Function Placement}

\subsection{Background}

When a specific SFC is deployed, it first instantiates the required VNFs as VNFIs, and then places these VNFIs in proper location of the network. 
Different VNF placement schemes can affect the network performance and placement cost. For example, as shown in Fig. 1(a), if only one VNFI for each VNF of SFC is placed in the network, 
the placement cost (approximatively the number of deployed VNFIs) is minimized, but the network performance is relatively low. 
SFC traffic throughput is equal to the available bottleneck bandwidth of path shown in Fig. 1(a), which may not satisfy users’ requirement. However, 
if the placement scheme as shown in Fig.e 1(b) is adopted, the network performance can be better (traffic throughput can be improved), but the placement cost also ascends. 
During the placement process, network operators usually hope to allocate minimized resources to each VNFI while satisfying the performance requirements. 
VNF placement optimization can also bring financial benefit for network operators (e.g. reduction of deployment and operational cost). 
Hence, academia and industry have paid more attention to identify optimal (or near optimal) VNF placement schemes for SFC.

\begin{figure}[!t]
  \centering
  \includegraphics[width=\linewidth]{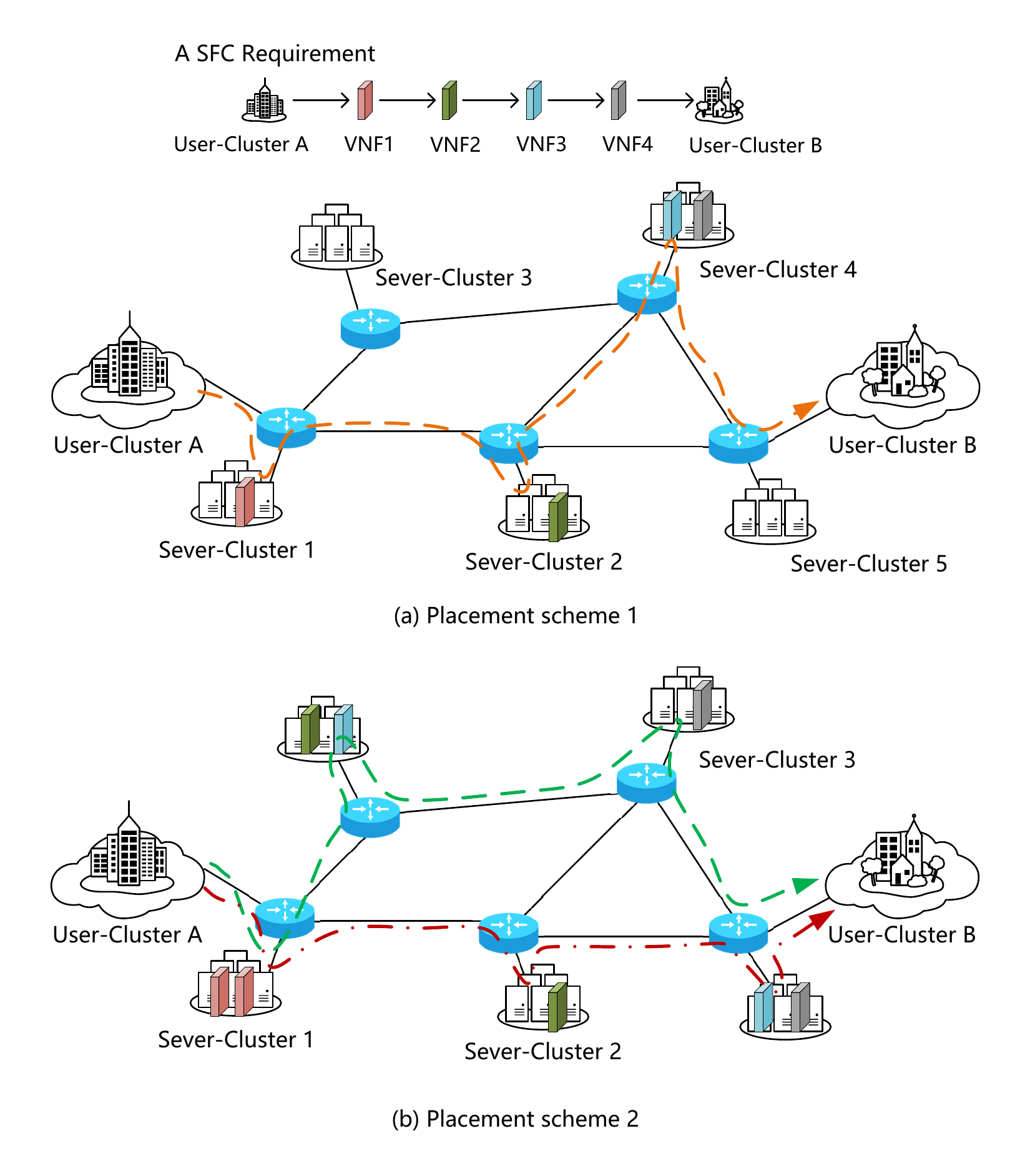}
  \caption{VNF placement in actual network example}
  \label{fig_1}
\end{figure}

\subsection{Existing Solutions}

n the current optimization solutions of VNF placement, the actual network is usually considered as a graph which includes a set of nodes and edges. 
The nodes are the abstract of forwarding devices in the network. Some of the nodes can connect with the server-clusters, and VNF can be deployed in these clusters. 
Each server-cluster has its own physical resources, containing CPU, memory, storage, etc. These resources should be allocated to the VNF as requirements. 
The edges in the graph represent the links between different nodes, and edges also have physical resource, mainly referring to link capacity. 
According to user requirements and resource constraints, the optimization solutions need to deploy VNFIs which are required by specific SFC in the graph, 
and then realize expected optimization goal. We illustrate some optimization solutions in detail as follows.

\subsubsection{Optimization Objective}
In general, the cost that physical devices use to run VNF is mainly considered. Ghaznavi et al. \cite{ghaznavi2017distributed} and Luizelli et al. 
\cite{luizelli2018optimizing} propose to use minimizing operational cost as the optimization objective. Particularly, Ghaznavi et al. \cite{ghaznavi2017distributed} aim to minimize 
the aggregate cost of allocating host and bandwidth resources. The host resources allocation cost is related to the resource demand for each VNF and the number of VNFIs running on host, 
and the bandwidth resources allocation cost is related to the volume of traffic on each link. Luizelli et al. \cite{luizelli2018optimizing} aim to minimize the virtual switching cost 
in physical devices, which is caused by software switching in order to steer traffic through VNFs of SFC.

\subsubsection{Optimization Problem Formulation}

Most optimization problems of VNF placement are modeled as Integer Programming problem \cite{luizelli2018optimizing} or Mixed Integer Programming (MIP) problem \cite{ghaznavi2017distributed}. 
Besides the optimization objective mentioned above, the problems also include the related resources and user demand constraints such as physical device capacity constraint, 
location constraint, link capacity constraint, throughput constraint and so on. These constraints are the boundary of VNF placement optimization problem, 
and they help to find optimal solution under specified conditions. Meanwhile, the computational complexity of solving the optimization problem also needs to be evaluated. Usually, 
the computational complexity is related to the number of nodes (namely physical devices that can run VNFs) in the network and the length of SFC (namely the number of VNFs in each SFC). 
More nodes or longer SFC means the computational complexity of solving process is higher.

\subsubsection{Algorithm Form} 

Some VNF placement optimization problems are proved as NP-hard problem (such as in research \cite{ghaznavi2017distributed}). 
That means it is difficult to realize fast solving for large-scale network. Therefore, some heuristic algorithms are proposed to realize fast solving. 
These heuristic algorithms include both classical algorithms (e.g. local search, greedy, etc.) and novel algorithms (e.g. bipartite graph matching \cite{luizelli2018optimizing}, etc.). 
For example, Ghaznavi et al. \cite{ghaznavi2017distributed} propose a local search heuristic solution called KARIZ. For a network topology (See Fig. 2(a)), 
it assumes each type of VNF in the SFC (See Fig. 2(b)) is deployed in a layer. Each layer contains a set of nodes in which the corresponding type of VNFIs can be installed (See Fig. 2(c)). 
The traffic can be routed layer by layer. During this process, the optimal routing between two layers is found by solving the minimum cost flow problem, 
and then the number of VNFIs in each layer is computed according to the allocated throughput. The algorithm repeats this process until the traffic has reached the last layer. 
Finally, the optimal result will be found (See Fig. 2(d)).

\begin{figure}[!t]
  \centering
  \includegraphics[width=\linewidth]{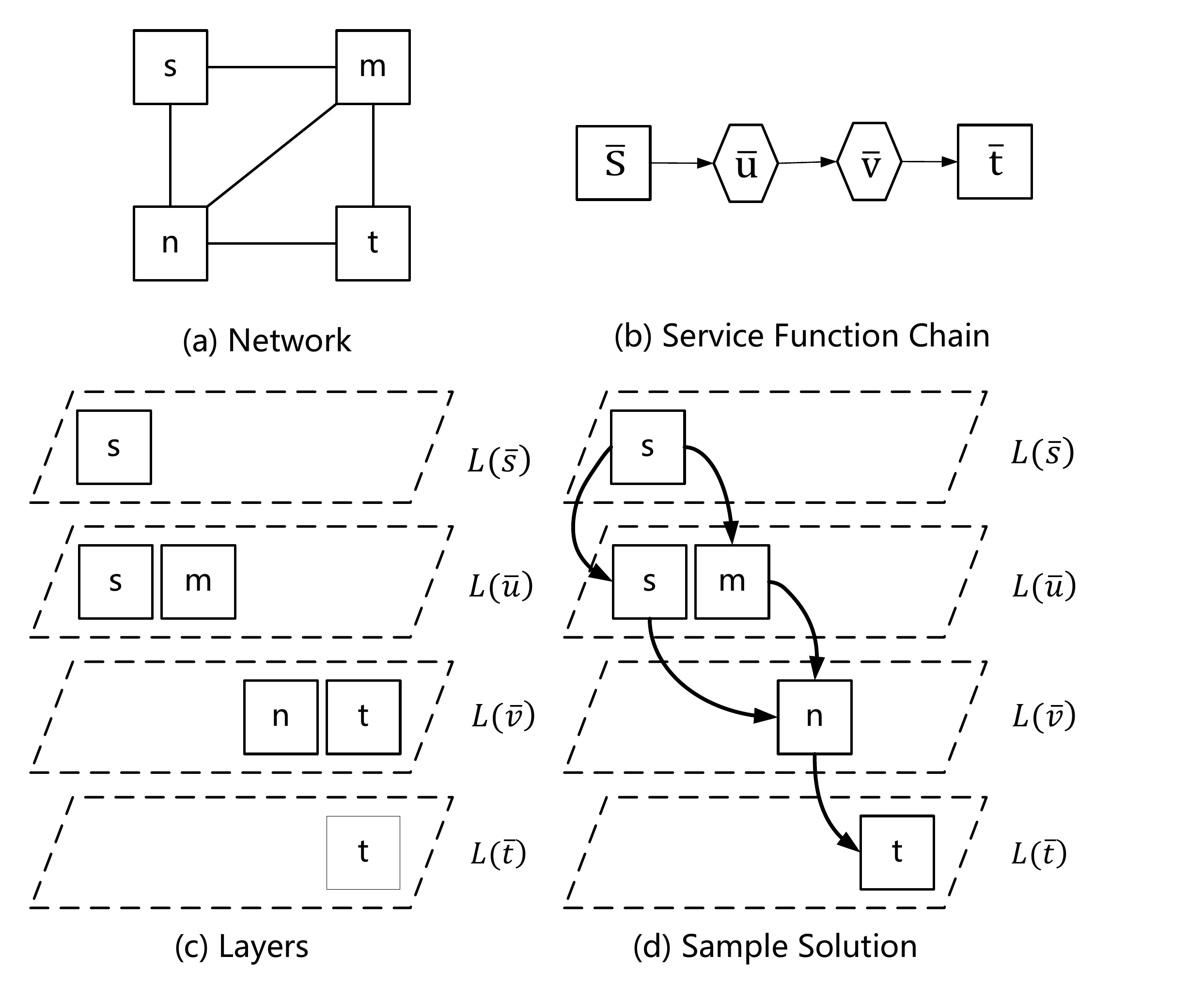}
  \caption{Layers example of KARIZ \cite{ghaznavi2017distributed}}
  \label{fig_2}
\end{figure}

\subsection{Summarization}

Existing VNF placement solutions mainly aim to minimize deployment cost and improve network performance. They model the optimization problem with integer 
linear programming and use heuristic algorithms to realize fast solving. Although there is some gap in term of accuracy between the heuristic algorithms and direct solving method, 
heuristic algorithms have advantage in computational complexity when solving large-scale network optimization problem.

\section{Service Function Chain Routing for Virtual Network Functions}

\subsection{Background}

Besides VNF placement, traffic routing also needs to be considered. The process of routing traffic requires to determine the forwarding path that traverses each VNF of SFC 
in specified order and consider the related network characteristics (such as link load, link transmission delay, etc.). The network operators usually wish to compute forwarding 
path efficiently and the routing cost could be minimized. In practice, traditional shortest path algorithm (like Dijkstra’s algorithm) can be helpful when computing forwarding path, 
but additional SFC constraints also need to be considered for satisfying user demands.

\subsection{Existing Solutions}

Similar to VNF placement optimization problem, SFC routing optimization problem also considers the actual network as a directed graph. 
The traffic should be transmitted from starting node to terminating node and pass through the VNFs of specified SFC. Meanwhile, 
the locations of these VNFs in the graph are assumed to be known in advance. The routing optimization solutions should calculate the shortest path with least cost and 
ensure the found paths are admissible.

\subsubsection{Optimization Objective}

The metric of SFC routing algorithm has many potential choices. It could be financial aspect (such as maintaining cost of forwarding devices, etc.) 
or network performance aspect (such as traffic propagation delay, user QoS demand, etc.). Existing optimization solutions usually aim to reduce the routing costs and improve the 
network performance like throughput \cite{sallam2018shortest}. For example, Dwaraki et al. \cite{dwaraki2016adaptive} use delay as the only metric for link communication and VNF processing, 
and then minimize the delay cost when calculating forwarding paths. The reason is that delay is an important consideration in many networks, 
and it can also be used to represent dynamic loads on network links and on VNF processing nodes. 

\subsubsection{Algorithm Form}

The SFC routing algorithms need to find a forwarding path that can transfer traffic from source to destination with least cost. Meanwhile, 
they also need to ensure the traffic can be processed by required network services. Dwaraki et al. \cite{dwaraki2016adaptive} propose an Adaptive Service Routing (ASR) algorithm that 
transforms the original network graph into a “layered graph” and uses conventional shortest-path algorithms to calculate forwarding paths. As shown in Fig. 3, 
each layer is a copy of the original graph. The number of layers is determined by the length of SFC (the SFC mentioned in Fig. 3 includes two VNFs, hence the figure contains three layers), 
and the order of layer (from top to bottom) implies the VNF order of SFC. Every layer is connected by the edges between nodes which can provide specific network function of SFC. 
The best path found by Dijkstra algorithm is the bold line as shown in Fig. 3, and the traffic flow is processed by both two VNFs of the SFC. 
Dijkstra algorithm uses communication delay and function processing delay as weight. It ensures the best path found by ASR has minimized delay, which is the optimal routing path. 

\begin{figure}[!t]
  \centering
  \includegraphics[width=2.5in]{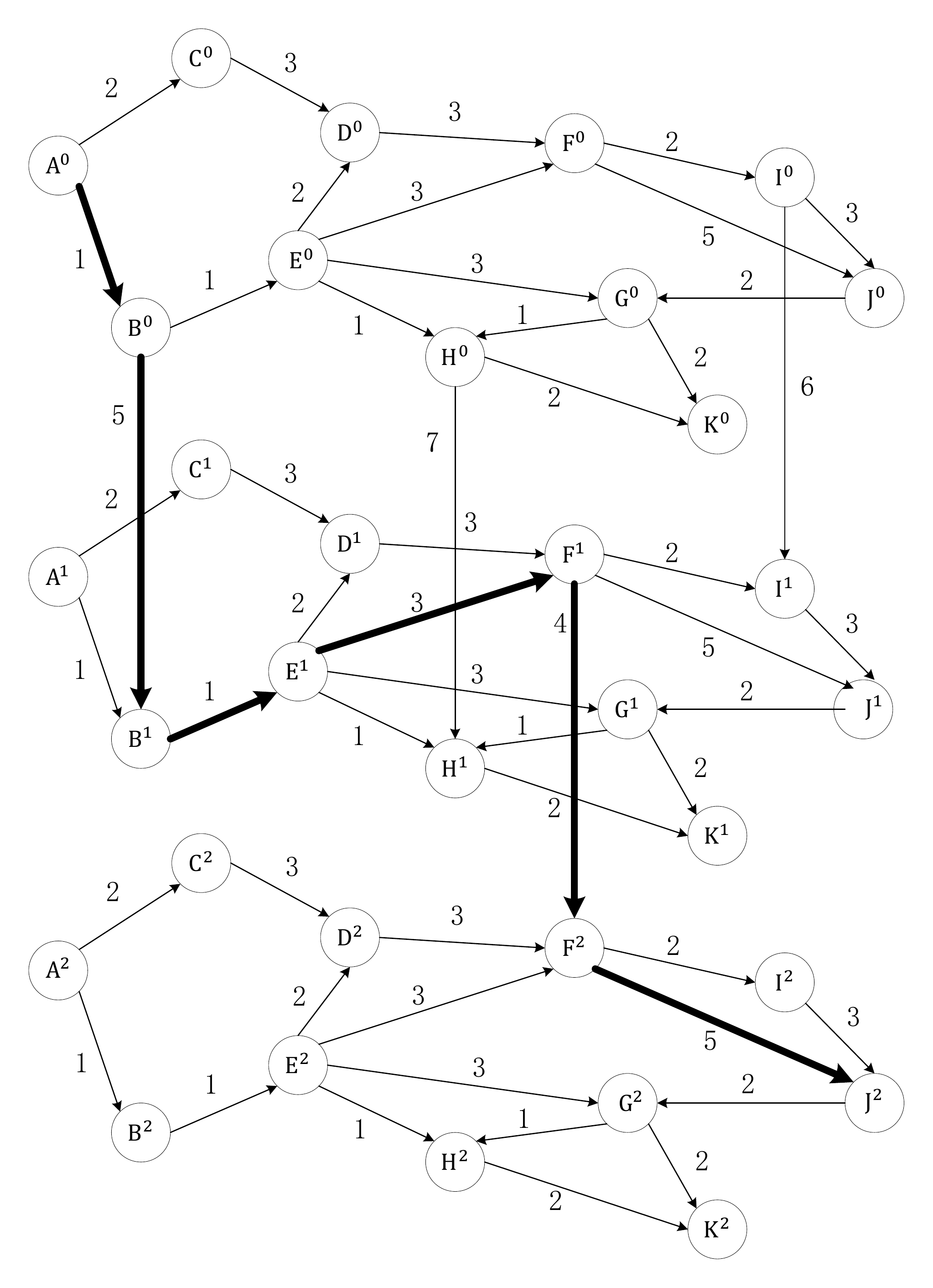}
  \caption{Layered graph for ASR shortest path algorithm \cite{dwaraki2016adaptive}}
  \label{fig_3}
\end{figure}

Sallam et al. \cite{sallam2018shortest} propose similar scheme which also constructs a new transformed graph and uses conventional shortest-path algorithms to compute 
SFC-constrained shortest path. The difference is that Sallam et al. \cite{sallam2018shortest} propose a pruning algorithm to simplify the constructed graph. 
It first constructs an initial graph (see Fig. 4(a)) that contains original node (white node) and several copies (gray node), and the number of copies also depends on the length of 
SFC (in Fig. 4, the example SFC contains two VNFs). The copy node is reachable if the path from one node (can be either original node or copy node) to itself can satisfy partial SFC. 
Then, it removes the nodes only have outgoing edges (except source node) and the nodes only have incoming edges (except destination node). 
After that, the pruned graph can be obtained (see Fig. 4(b)). This difference can help to reduce the computational time when using shortest path algorithm compared with ASR algorithm. 

\begin{figure}[!t]
  \centering
  \includegraphics[width=2.5in]{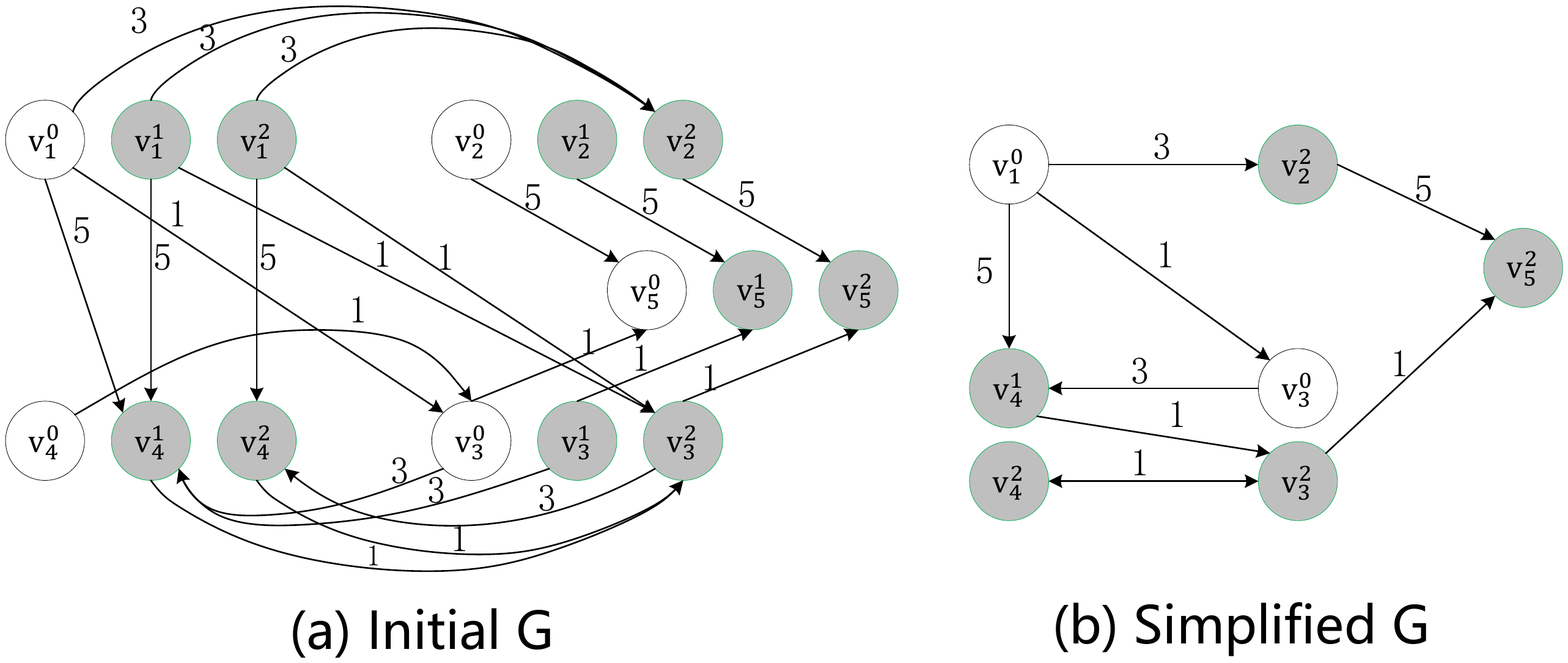}
  \caption{Initial and pruned graph for SFC-constrained shortest path algorithm \cite{sallam2018shortest}}
  \label{fig_4}
\end{figure}

\subsection{Summarization}

The major objectives of existing SFC routing solutions are routing traffic with least cost and meeting SFC demands. 
They usually transform the routing problem into shortest path problem with SFC constraints, and then use conventional shortest path algorithms to solve this problem. 
The metric used to select optimal path can be various, mainly depending on the choice of network operators. Meanwhile, 
the efficiency of computing SFC-constrained shortest path can also be guaranteed in large-scale network.

\section{Joint Optimization of VNF Placement and SFC routing}

\subsection{Background}

When VNF placement and SFC routing optimization problems are considered jointly, there cloud be a conflict between these two problems. 
For example, as shown in Fig. 5(a) and 5(b) (here we use the topology similar to [9]), there are three traffic requests T1 (from node 3 to 11), T2 (from node 11 to 1) and T3 
(from node 10 to 5) demand SFC1 composed of VNF1, VNF2 and VNF3 (the order of VNFs is VNF1-VNF2-VNF3). In Fig. 5(a), if there is only one instance of SFC1 in the network, 
traffic flow T2 and T3 have to be routed over longer path, which causes more routing cost. However, if we deploy two SFC1 instances in the network, as shown in Fig. 5(b), 
the routing cost can be reduced due to using shorter forwarding paths. This example implies that optimizing VNF placement alone by instantiating fewer VNFIs may cause the traffic 
routing cost increasing. Whereas, if SFC routing optimization is considered preferentially, the additional VNF placement cost may be introduced, because more VNFIs are required to satisfy 
abundant traffic demand in today’s network environment. Hence, joint optimization of VNF placement and SFC routing is necessary to find a trade-off optimal SFC deployment scheme.

\begin{figure*}[!t]
  \centering
  \includegraphics[width=\linewidth]{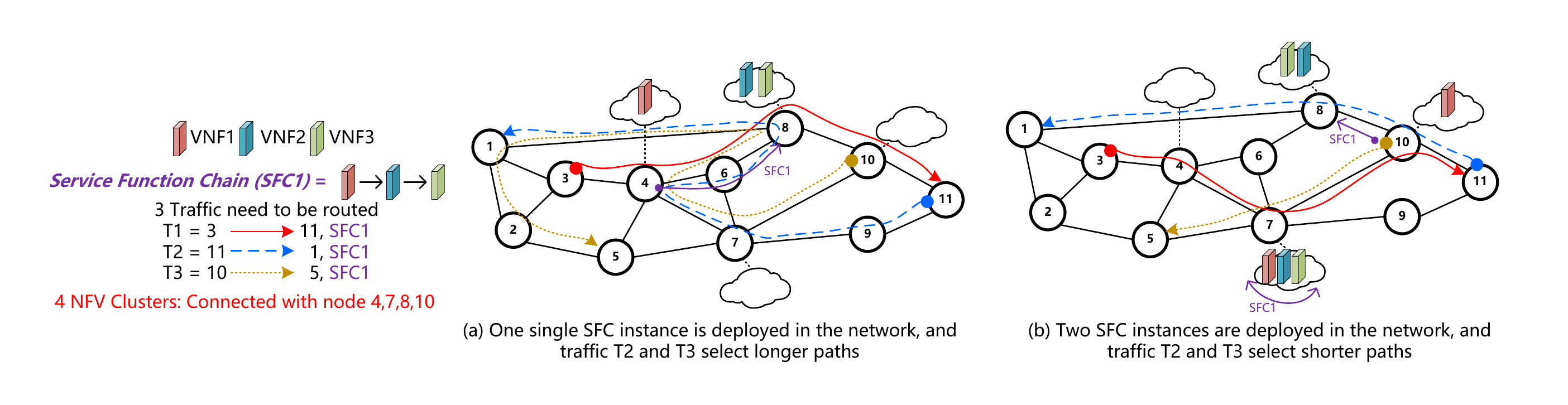}
  \caption{Conflict between VNF placement and SFC routing}
  \label{fig_5}
\end{figure*}

\subsection{Existing Solutions}

Joint optimization solutions should deploy required VNFs of SFC properly, which means the deployment scheme can achieve high resource utilization or minimize the resources that need 
to be allocated with VNFs. Meanwhile, user traffic flow should also be routed through specified VNFs with QoS requirements. Besides these tasks, 
some solutions also consider the migration of VNFIs in response to the variation of user demand or network situation. Next, we will introduce some existing joint optimization 
schemes for VNF placement and SFC routing.

\subsubsection{Optimization Objective}

The objective of VNF placement and SFC routing joint optimization can be diverse. Some joint optimization solutions usually combine the VNF placement and SFC routing optimization objectives 
together. For example, Addis et al. \cite{addis2015virtual} propose using minimization of the maximum link utilization as network-level optimization objective, 
and minimization of allocated computing resources as VNFI-level optimization objective. And Zhang et al. \cite{zhang2017joint} use maximizing the average resource utilization of each computing 
node and minimizing the average response latency of traffic scheduling as optimization objective. Since most existing optimization solutions belong to multi-objective optimization, 
they usually use weighted sum approach to represent the joint optimization objective.

On the other hand, some solutions do not explicitly represent the VNF placement and SFC routing optimization objectives mentioned above. For example, Gupta et al. \cite{gupta2018scalable} aim to 
minimize bandwidth consumption by instantiating proper number of VNFs and selecting shortest path for routing traffic. Similarly, Guo et al. \cite{guo2018joint} and Qu et 
al. \cite{qu2018reliability} select maximizing resource utilization as the main optimization objective. In addition, considering reconstruction for variation of user demand or network situation, 
Tajiki et al. \cite{tajiki2018joint} and Eramo et al. \cite{eramo2017approach} take minimization of energy consumption and reconfiguration cost into account as optimization objective. 
Meanwhile, Tajiki et al. \cite{tajiki2018joint} aim to minimize energy consumption by reducing the number of hops that the flow needs to traverse. And Eramo et al. \cite{eramo2017approach} also 
try to minimize the rejected bandwidth for better quality of service. 

\subsubsection{Optimization Problem Formulation}

The type of optimization problem formulation mainly depends on the optimization objective. If the optimization objective is the combination of VNF placement and SFC routing optimization objectives, 
the joint optimization problem is usually modeled as Mixed Integer Linear Programming (MILP) problem \cite{addis2015virtual,zhang2017joint}. The reason is besides integer variables 
(like physical resources capacity), some SFC routing optimization solutions may involve real variables (like link delay). For instance, Addis et al. \cite{addis2015virtual} use RAM and CPU to 
express the VNF computing resource consumption and total transmission latency is introduced by VNF forwarding. Each VNFI node has their own resource capacity constraint and 
latency also has maximum bound. The optimization solution needs to minimize the maximum network link utilization and number of cores (CPU) used by the instantiated VNFs within these constraints.

By contrast, if the optimization objective does not involve real variables, the optimization solutions usually use ILP to model the optimization 
problem \cite{gupta2018scalable,tajiki2018joint,eramo2017approach,qu2018reliability}. For example, Gupta et al. \cite{gupta2018scalable} aim to minimize bandwidth consumed. 
It precomputes the potential set of configurations for SFC and uses them as input for the ILP model. The ILP model can select the best configuration based on related constraints, 
and then compute the forwarding path for user traffic. The constraints mainly include the number of physical nodes allowed to deploy VNFs, sufficient number of CPU for running VNFs, 
link capacity and so on.

\subsubsection{Algorithm Form}

Since the joint optimization problems of VNF placement and SFC routing are basically NP-hard, most solutions propose corresponding heuristic algorithms to realize rapid solving. 
The details of each heuristic algorithm can be different according to the specific optimization problems. But the main idea of these heuristic algorithms is similar. 
They all rely on related network operational experience, leverage constraint relaxation, iteration and other methods to achieve the trade-off between optimality gap and computational complexity, 
and then find the result that is close to the optimal solution. However, the results solved by heuristic algorithm are usually near-optimal and the gap 
between near-optimal and optimal solutions cannot be estimated. Some typical examples of heuristic algorithms are presented as follows.

Heuristic algorithms usually obtain the near optimal solution through continuous iteration. For example, in Addis et al. research \cite{addis2015virtual}, there are two competitive optimization 
objectives: minimizing total virtualization cost (first objective) and minimizing maximum link utilization (second objective). Because this research prefers to improve user service quality, 
it first finds the best result according to the first objective, and then increases the value found in first objective step by step until the desired cost level of the second objective is found. 
Finally, the optimal VNF deployment and traffic routing policy can be determined. And Eramo et al. \cite{eramo2017approach} propose an efficient heuristic algorithm which sorts the offered SFCs 
in bandwidth decreasing order, and then determines the mapping (contains VNF placement and traffic routing) scheme for each SFC in sequence. When all SFCs have been traversed, 
the algorithm can find the optimal SFC configuration. For maximization of resource utilization, Qu et al. \cite{qu2018reliability} propose a bi-directional search methodology. 
It uses greedily search and shortest path routing to select the best physical machines that have enough computing resources to run VNFIs of the SFC. The algorithm executes both forward search 
(from source node of traffic) and backward search (from destination node of traffic). Backward search can help to improve the result found by the forward search. This method can avoid the 
algorithm trapping into local optimum.

\begin{table*}[!t]
  \caption{Comparison for different optimization solutions}
  \label{table_1}
  \centering
  \begin{tabular}{|m|m|m|m|m|m|m|m|}
  \hline
  \multicolumn{1}{|m{1.5cm}<{\centering}|}{Optimization Type} & \multicolumn{1}{|m{1cm}<{\centering}|}{Specific Works} & \multicolumn{1}{|m{1.5cm}<{\centering}|}{Optimization Objective} 
  & \multicolumn{1}{|m{1.5cm}<{\centering}|}{Formulation Type} & \multicolumn{1}{|m{2cm}<{\centering}|}{Algorithm Type} & \multicolumn{1}{|m{2.5cm}<{\centering}|}{Complexity} 
  & \multicolumn{1}{|m{3cm}<{\centering}|}{Strength} & \multicolumn{1}{|m{3cm}<{\centering}|}{Weakness}\\
  \hline
  \multicolumn{1}{|m{1.5cm}<{\centering}|}{\multirow{2}{1.5cm}[-0.5cm]{\centering VNF Placement}}
  & \multicolumn{1}{|m{1cm}<{\centering}|}{KARIZ \cite{ghaznavi2017distributed}} & \multicolumn{1}{|m{1.5cm}<{\centering}|}{MIN deployment cost} & \multicolumn{1}{|m{1.5cm}<{\centering}|}{MIP} 
  & \multicolumn{1}{|m{2cm}<{\centering}|}{Heuristic} &  \multicolumn{1}{|m{2.5cm}<{\centering}|}{$O(VN^{2}\log N(E+\log N^{2}))$} & \multicolumn{1}{|m{3cm}<{\centering}|}{Well optimize CPU cost during VNF placement, and time complexity is reasonable} 
  & \multicolumn{1}{|m{3cm}<{\centering}|}{The lower bound of algorithm cannot be guaranteed}\\
  \cline{2-8}
  \multicolumn{1}{|m{1.5cm}<{\centering}|}{}
  & \multicolumn{1}{|m{1cm}<{\centering}|}{OCM \cite{luizelli2018optimizing}} & \multicolumn{1}{|m{1.5cm}<{\centering}|}{MIN (switching) cost} & \multicolumn{1}{|m{1.5cm}<{\centering}|}{N/A} 
  & \multicolumn{1}{|m{2cm}<{\centering}|}{Heuristic} & \multicolumn{1}{|m{2.5cm}<{\centering}|}{$O(V^{V}(V+N)V^{2}N^{2})$} & \multicolumn{1}{|m{3cm}<{\centering}|}{Optimize internal switching CPU cost to improve network utilization} 
  & \multicolumn{1}{|m{3cm}<{\centering}|}{Time complexity of algorithm is affected by SFC length obviously}\\
  \hline
  \multicolumn{1}{|m{1.5cm}<{\centering}|}{\multirow{2}{1.5cm}[-0.5cm]{\centering SFC Routing}}
  & \multicolumn{1}{|m{1cm}<{\centering}|}{ASR \cite{dwaraki2016adaptive}} & \multicolumn{1}{|m{1.5cm}<{\centering}|}{MIN total routing delay} & \multicolumn{1}{|m{1.5cm}<{\centering}|}{N/A} 
  & \multicolumn{1}{|m{2cm}<{\centering}|}{N/A} &  \multicolumn{1}{|m{2.5cm}<{\centering}|}{$O((E+N)\log N)$} & \multicolumn{1}{|m{3cm}<{\centering}|}{Use shortest path algorithms to simplify traffic routing optimization in layered graph} 
  & \multicolumn{1}{|m{3cm}<{\centering}|}{The large size of layered graph may affect algorithm run time}\\
  \cline{2-8}
  \multicolumn{1}{|m{1.5cm}<{\centering}|}{}
  & \multicolumn{1}{|m{1cm}<{\centering}|}{SCSP \cite{sallam2018shortest}} & \multicolumn{1}{|m{1.5cm}<{\centering}|}{MIN routing cost} & \multicolumn{1}{|m{1.5cm}<{\centering}|}{N/A} 
  & \multicolumn{1}{|m{2cm}<{\centering}|}{N/A} & \multicolumn{1}{|m{2.5cm}<{\centering}|}{$O(E\log N)$} & \multicolumn{1}{|m{3cm}<{\centering}|}{Simplify the layered graph in \cite{dwaraki2016adaptive}, and improve shortest path algorithm efficiency} 
  & \multicolumn{1}{|m{3cm}<{\centering}|}{Ignore the VNF execute cost on network node}\\
  \hline
  \multicolumn{1}{|m{1.5cm}<{\centering}|}{\multirow{7}{1.5cm}[-3cm]{\centering Joint optimization of VNF placement and SFC routing}}
  & \multicolumn{1}{|m{1cm}<{\centering}|}{VNF-PR \cite{addis2015virtual}} & \multicolumn{1}{|m{1.5cm}<{\centering}|}{MIN maximum link utilization and host cores number} & \multicolumn{1}{|m{1.5cm}<{\centering}|}{MILP} 
  & \multicolumn{1}{|m{2cm}<{\centering}|}{Heuristic} &  \multicolumn{1}{|m{2.5cm}<{\centering}|}{Not evaluated} & \multicolumn{1}{|m{3cm}<{\centering}|}{Acceptable execution time for large scale optimization problem} 
  & \multicolumn{1}{|m{3cm}<{\centering}|}{No specific time complexity}\\
  \cline{2-8}
  \multicolumn{1}{|m{1.5cm}<{\centering}|}{}
  & \multicolumn{1}{|m{1cm}<{\centering}|}{BFDSU\& RCKK \cite{zhang2017joint}} & \multicolumn{1}{|m{1.5cm}<{\centering}|}{MAX resource utilization and MIN average latency} & \multicolumn{1}{|m{1.5cm}<{\centering}|}{MILP} 
  & \multicolumn{1}{|m{2cm}<{\centering}|}{Approximation (BFDSU) Heuristic (RCKK)} & \multicolumn{1}{|m{2.5cm}<{\centering}|}{$O(\log V+N\log N)$ (BFDSU) $O(NV\log V)$ (RCKK)} & \multicolumn{1}{|m{3cm}<{\centering}|}{Worst-case performance bound of algorithm (BFDSU) performance has been theoretically proved} 
  & \multicolumn{1}{|m{3cm}<{\centering}|}{Optimization effect of request scheduling is not obvious}\\
  \cline{2-8}
  \multicolumn{1}{|m{1.5cm}<{\centering}|}{}
  & \multicolumn{1}{|m{1cm}<{\centering}|}{MWUA \cite{guo2018joint}} & \multicolumn{1}{|m{1.5cm}<{\centering}|}{MAX overall resource utility} & \multicolumn{1}{|m{1.5cm}<{\centering}|}{ILP} 
  & \multicolumn{1}{|m{2cm}<{\centering}|}{Approximation} & \multicolumn{1}{|m{2.5cm}<{\centering}|}{Not evaluated} & \multicolumn{1}{|m{3cm}<{\centering}|}{Upper and lower bound of algorithm performance has been theoretically proved} 
  & \multicolumn{1}{|m{3cm}<{\centering}|}{Problem parameters is coarse-grained}\\
  \cline{2-8}
  \multicolumn{1}{|m{1.5cm}<{\centering}|}{}
  & \multicolumn{1}{|m{1cm}<{\centering}|}{SPTG\& CG-ILP \cite{gupta2018scalable}} & \multicolumn{1}{|m{1.5cm}<{\centering}|}{MIN bandwidth consumed} & \multicolumn{1}{|m{1.5cm}<{\centering}|}{ILP} 
  & \multicolumn{1}{|m{2cm}<{\centering}|}{Heuristic} & \multicolumn{1}{|m{2.5cm}<{\centering}|}{High in large scale network (not formulized)} & \multicolumn{1}{|m{3cm}<{\centering}|}{Well optimize the bandwidth consumed in WAN scenario with heavy traffic} 
  & \multicolumn{1}{|m{3cm}<{\centering}|}{Run time of CG-ILP is not acceptable in large scale network}\\
  \cline{2-8}
  \multicolumn{1}{|m{1.5cm}<{\centering}|}{}
  & \multicolumn{1}{|m{1cm}<{\centering}|}{NSF \cite{tajiki2018joint}} & \multicolumn{1}{|m{1.5cm}<{\centering}|}{MIN energy consumption} & \multicolumn{1}{|m{1.5cm}<{\centering}|}{ILP} 
  & \multicolumn{1}{|m{2cm}<{\centering}|}{Heuristic} & \multicolumn{1}{|m{2.5cm}<{\centering}|}{$O(\psi (N\log N+E))$ ($\psi$ is parameter)} & \multicolumn{1}{|m{3cm}<{\centering}|}{Novel solutions for energy-aware management of network traffic, low execution time} 
  & \multicolumn{1}{|m{3cm}<{\centering}|}{Optimality gap is evaluated by experiment, lack of theoretical proof}\\
  \cline{2-8}
  \multicolumn{1}{|m{1.5cm}<{\centering}|}{}
  & \multicolumn{1}{|m{1cm}<{\centering}|}{MASB \& VMMPC \& RLACM \cite{eramo2017approach}} & \multicolumn{1}{|m{1.5cm}<{\centering}|}{MIN fraction of total rejected and migration cos} & \multicolumn{1}{|m{1.5cm}<{\centering}|}{ILP} 
  & \multicolumn{1}{|m{2cm}<{\centering}|}{Heuristic} & \multicolumn{1}{|m{2.5cm}<{\centering}|}{$O(V(V+N_{s})N_{l}\log (N_{s}+N_{n}))$ (MASB) $O(N_{s}^{3}N_{l}\log (N_{s}+N_{n}))$ (VMMPC)} & \multicolumn{1}{|m{3cm}<{\centering}|}{Well optimize migration cost, take advantage of Viterbi algorithm for migration policy} 
  & \multicolumn{1}{|m{3cm}<{\centering}|}{Lack of theoretical evaluation for optimality gap}\\
  \cline{2-8}
  \multicolumn{1}{|m{1.5cm}<{\centering}|}{}
  & \multicolumn{1}{|m{1cm}<{\centering}|}{REACH \cite{qu2018reliability}} & \multicolumn{1}{|m{1.5cm}<{\centering}|}{MAX network resources utility} & \multicolumn{1}{|m{1.5cm}<{\centering}|}{ILP} 
  & \multicolumn{1}{|m{2cm}<{\centering}|}{Heuristic} & \multicolumn{1}{|m{2.5cm}<{\centering}|}{$O(N^{2})$} & \multicolumn{1}{|m{3cm}<{\centering}|}{Use bi-direction search to avoid local optima} 
  & \multicolumn{1}{|m{3cm}<{\centering}|}{Lack of evaluation for optimality gap}\\
  \hline
  \end{tabular}
  \begin{tablenotes}
    \footnotesize
      \item[1] N/A means the solution does not give out the type of optimization problem formulation or algorithm.
      \item[2] MIN means minimize, MAX means maximize.
      \item[3] In complexity domain, $N$ means the number of physical devices which can hold VNFI, $E$ means the number of links, $V$ means the VNF number of SFC. 
      Particularly, in \cite{eramo2017approach}, $N_{s}$,$N_{n}$,$N_{l}$ means the number of servers, switch and links of network respectively.
  \end{tablenotes}
  \end{table*}

  Moreover, some existing solutions propose approximation algorithms to solve the joint optimization problem. For example, Zhang et al. \cite{zhang2017joint} design a priority-driven 
  weighted algorithm to find near optimal solution. The algorithm calculates the probability of placing VNF at a physical device by its reciprocal of RST 
  (RST refers to remaining resource capacity of the physical device), and then places the VNF with the maximum probability for maximizing network resource utilization. Similarly, 
  Guo et al. \cite{guo2018joint} propose a multiplicative weight update algorithm. It first formulates the dual of the original optimization problem, 
  and then introduces dual variable for user traffic flow and weight variable for related physical resources. The algorithm will assign the SFC configuration for the adopted flow, 
  and the weight variable will also be updated. The algorithm will be executed until all arrival flow is traversed. Unlike heuristic algorithm, 
  approximation algorithm can guarantee the gap between the result solved by itself and optimum within bounds.

\subsection{Summarization}

VNF placement and SFC routing joint optimization solutions have the optimization objectives in both VNF-level (mainly consider deployment cost, resource usage, etc.) 
and routing-level (mainly consider link utilization, delay, etc.). Because of the conflict between these two levels, the optimization solutions need to balance the objectives of 
VNF-level and routing-level according to the requirements of network operators and users. Furthermore, in order to realize fast solving in large-scale network, 
these solutions propose different heuristic algorithms or approximation algorithms to exchange the accuracy of optimization results for lower computational complexity. 

\section{Comparison for Different Optimization Solutions}

In this section, we will compare the different SFC placement and routing optimization solutions mentioned above. They are compared based on the optimization type, 
the objective of optimization problem, the formulation that used to model the optimization problem, algorithm type, algorithm complexity, algorithm strength and weakness. 
The details of the comparison are shown in table 1.


\section{Future research prospects}

At present, a lot of research has proposed corresponding solutions which optimize VNF deployment and traffic routing scheme for better performance. However, 
the user demands can usually be variable in real-time. If the SFC configurations cannot be adjusted to accommodate the variations, the network performance may decline 
(such as resources utilization decreasing, response latency increasing, etc.). Actually, most existing solutions don’t consider this problem. Based on the real needs, 
SFC elastic scaling (or dynamic adjustment) is a good research direction. Two main kinds of elastic scaling approaches are shown as following.

\subsection{Auto-scaling based on threshold}

Adel et al. \cite{toosi2019elasticsfc} propose a dynamic auto-scaling algorithm called ElasticSFC to allocate or release VNF and bandwidth resource. 
The scaling decision is made depending on whether the CPU utilization of physical host or bandwidth consumption is higher than upper bound 
(or less than lower bound). However, scaling approaches based on threshold are reactive to adjust the SFC deployment scheme or routing policy, 
namely adjust SFC configurations after variations have happened (may have happened for a while). This may not be the best solution.

\subsection{Auto-scaling based on demand prediction}

Demand prediction can be used to determine the extent of scaling VNF instances dynamically and the forwarding paths of flow can also be adjusted according to the variants of VNFIs. 
Some online learning methods have been used in recent researches. For example, Fei et al. \cite{fei2018adaptive} propose an online-learning method called follow-the-regularized-leader (FTRL) 
for upcoming user flows prediction. It can directly predict the flow rates of SFC and help to determine the scaling strategy of VNFIs for minimizing deployment cost.

On the other hand, machine learning technology has attracted a lot of attention in the field of networking. It can be helpful in traffic classification, online traffic scheduling, 
routing decisions and so on. There are some solutions using deep learning technology in VNF selecting and chaining problem. Instead of traditional heuristic algorithms, 
they use deep learning techniques to solve optimization problems \cite{pei2018virtual}. These methods can yield time efficiency and scalability benefit. 
Hence, combining machine learning technology with SFC placement and routing optimization problem can be another expected research direction in the future.

\section{Conclusion}

In this article, we first introduce VNF placement and SFC routing optimization problems independently. Then the joint optimization problem of VNF placement and SFC routing is introduced. 
For each kind of optimization problem, we describe the problem background, optimization objective, optimization problem formulation and algorithm form in details. 
Moreover, we also summarize and compare recent existing solutions, and then propose the future research prospects of SFC placement and routing problem.

\ifCLASSOPTIONcaptionsoff
  \newpage
\fi



%

%

\begin{IEEEbiographynophoto}{Weihan Chen}
  (chenwh18@mails.tsinghua.edu.cn) is currently working toward the Ph.D. degree with the department of computer science and technology, Tsinghua University. 
  He received the B.E. degree in telecommunications engineering and management and M.E. degree in software engineering from Beijing University of Posts and Telecommunications, China. 
  His current research interests include network function virtualization, software defined networking and traffic engineering.
\end{IEEEbiographynophoto}
 
\begin{IEEEbiographynophoto}{Xia Yin}
  [SM] (yxia@tsinghua.edu.cn) received the B.E., M.E. and Ph.D. degrees in computer science from Tsinghua University in 1995, 1997 and 2000 respectively. 
  She is a Full Professor in Department of Computer Science and Technology at Tsinghua University. Her research interests include future Internet architecture, formal method, 
  protocol testing and large-scale Internet routing.
\end{IEEEbiographynophoto}
\vspace{-500 pt} 
\begin{IEEEbiographynophoto}{Zhiliang Wang}
  [M] (wzl@cernet.edu.cn) received the B.E., M.E. and Ph.D. degrees in computer science from Tsinghua University, China in 2001, 2003 and 2006 respectively. 
  Currently he is an Associate Professor in the Institute for Network Sciences and Cyberspace at Tsinghua University. His research interests include formal methods and protocol testing, 
  next generation Internet, network measurement.
\end{IEEEbiographynophoto}
\vspace{-500 pt}  
\begin{IEEEbiographynophoto}{Xingang Shi}
  [M] (shixg@cernet.edu.cn) received the B.S. degree from Tsinghua University and the PhD degree from The Chinese University of Hong Kong.
   He is now working in the Institute for Network Sciences and Cyberspace at Tsinghua University. His research interests include network measurement and routing protocols.
\end{IEEEbiographynophoto}







\end{document}